%% file: main.tex
\documentclass[9pt,article,a4paper,twoside,onecolumn]{memoir}

\usepackage{fixltx2e}
\usepackage{hyperref}
\usepackage{breakurl}
\usepackage{multicol}
\usepackage[latin1]{inputenc}
\usepackage[T1]{fontenc}
\usepackage{mathtools}
\usepackage{color,graphicx,datetime}
\usepackage[scaled=0.92]{helvet}
\usepackage{mathptmx}
\usepackage{ragged2e}
\usepackage[sectionbib,sort&compress]{natbib}
\providecommand{\hslash}{\hbar}
\providecommand{\mbf}{\mathbf}
\renewcommand*{\vec}[1]{\mbf{#1}}
\newcommand*{\B}{\vec{B}}
\newcommand*{\E}{\vec{E}}
\newcommand*{\J}{\vec{J}}
\renewcommand*{\S}{\vec{S}}
\newcommand*{\x}{\vec{x}}
\newcommand*{\0}{\vec{0}}
\newcommand*{\cross}{{\times}}
\newcommand*{\df}[1]{\mathrm{d}#1}
\newcommand*{\dV}{\df{^3}\!x}
\newcommand*{\intV}[2]{\int_{V{#1}}\!\dV{#1}\,{#2}}

\renewcommand{\footnoterule}{%
 \kern-3pt%
 \hrule width \textwidth%
 \kern 2.6pt
}
\thanksmarkseries{arabic}
\setsecnumdepth{chapter}
\setsecheadstyle{\large\rmfamily\raggedright}
\setbeforesecskip{-3.0ex plus -1ex minus -.2ex}
\setaftersecskip{.4ex plus .2ex minus .1ex}
\setbeforesubsecskip{-3.0ex plus -1ex minus -.2ex}
\setaftersubsecskip{.2ex plus .05ex minus .02ex}

\captionnamefont{\small\bfseries\rmfamily}
\captiontitlefont{\small\rmfamily}
\captiondelim{\,\textbf{|}\,}
\captionstyle[\centering]{\raggedright}

\newenvironment{firstparagraph}{%
 \setlength{\parindent}{0in}%
 \setlength{\parskip}{0in}%
 \bfseries%
}{\par}

\newenvironment{methods}{%
 \section*{METHODS SUMMARY}%
 \small
}{}

\newenvironment{acknowledgements}{%
 \setparahook{\small\sffamily}%
 \setparaheadstyle{\small\sffamily\bfseries}%
 \paragraph{Acknowledgements}
}{\normalsize}

\pretitle{%
 \noindent
 \rule{\textwidth}{0.2pt}\\[-1.8ex]
 \rule{\textwidth}{0.1pt}\\[-1.8ex]
 \rule{\textwidth}{0.2pt}\\[-1.5ex]
 \begin{flushleft}\rmfamily\bfseries
 \fontsize{24}{24}\selectfont
}
\title{%
 Radio beam vorticity and orbital angular momentum
}
\posttitle{%
\par\end{flushleft}\vspace*{5ex}
}
\preauthor{\rmfamily\large\noindent}
\author{%
Bo Thidé%
\thanks{%
 Swedish Institute of Space Physics,
 Box 537, SE-75121 Uppsala, Sweden, EU.
 },
Fabrizio Tamburini%
\thanks{%
 Department of Astronomy,
 University of Padova,
 vicolo dell'Osservatorio 3,
 IT-33122 Padova, Italy, EU.
 },
Elettra Mari%
\thanks{%
 CISAS, Centro Interdipartimentale di Studi e Attività Spaziali G. Colombo,
 University of Padova,
 Via Venezia 15, IT-35131 Padova, Italy, EU.
},
Filippo Romanato%
 \thanks{%
  Department of Physics,
  University of Padova,
  via Marzolo,
  IT-35131, Padova, Italy, EU.
 }%
 \thanks{%
  LaNN, Laboratory for Nanofabrication of Nanodevices,
  Venetonanotech, via Stati Uniti 4,
  IT-35100 Padova, Italy, EU.
 }
\&
Cesare Barbieri%
\thanksmark{2}
}
\postauthor{\normalfont}

\predate{}
\date{}
\postdate{\vspace*{2.5ex}}

\begin{document}

\maketitle

\begin{multicols}{2}

\begin{firstparagraph}

It has been known for a century that electromagnetic             
fields can transport not only energy and linear momentum but            
also angular momentum\cite{Poynting:PRSL:1909,Abraham:PZ:1914}.         
However, it was not until twenty years ago, with the discovery 
in laser optics of experimental techniques for the generation,          
detection and manipulation of photons in well-defined, pure orbital     
angular momentum (OAM) states\cite{Beijersbergen&al:OC:1994}, that       
twisted light and its pertinent optical vorticity and phase            
singularities began to come into widespread use in science and         
technology\cite{Molina-Terriza&al:NP:2007,Franke-Arnold&al:LPR:2008,    
Torres&Torner:Book:2011}. We have now shown experimentally how 
OAM and vorticity can be readily imparted onto radio beams. Our results      
extend those of earlier experiments on angular momentum and vorticity   
in radio\cite{Carrara:N:1949,Carrara:NC:1949,diFrancia:NC:1957,         
Allen:AJP:1966,Carusotto&al:NC:1968,Hajnal:PRSLA:1990,                  
Kristensen&Woerdman:PRL:1994,Courtial&al:PRL:1998,                      
Jiang&al:Inproceedings:2009} in that we used a single antenna and       
reflector to directly generate twisted radio beams and verified that    
their topological properties agree with theoretical predictions. This   
opens the possibility to work with photon OAM at frequencies low enough 
to allow the use of antennas and digital signal processing, thus enabling      
software controlled experimentation also with first-order quantities,   
and not only second (and higher) order quantities as in optics-type     
experiments\cite{Thide&al:PRL:2007}. Since the OAM state space is 
infinite, our findings provide new tools for achieving high   
efficiency in radio communications and radar technology.                

\end{firstparagraph}                                                    

The total angular momentum around the origin ${\x=\0}$ carried by an
electromagnetic field $(\E,\B)$ in a volume $V$ of free space
is\cite{Schwinger&al:Book:1998,Jackson:Book:1999,Thide:Book:2011}
${
\J = \varepsilon_0\intV{}{\,\x\cross(\E\cross\B)}
}
$
where $\varepsilon_0$ is the free-space dielectric permittivity.        
For a beam where the fields fall off sufficiently         
rapidly with distance from the beam axis, the total angular             
momentum can be written $\J=\S+\vec{L}$, where the spin angular              
momentum $\S$ is associated with the two states of wave                 
polarisation and $\vec{L}$ is the orbital angular momentum (OAM).

Optical vortices are phase defects\cite{Nye&Berry:PRSLA:1974}           
embedded in light beams that carry OAM                                  
\cite{Allen&al:PRA:1992,Padgett&al:AJP:1996,Padgett&Allen:OC:1995}.     
Pure OAM state beams are characterized by a quantized topological       
charge $\ell$, which can take any integer value, corresponding          
to an OAM of $\ell\hslash$ carried by each photon of the                
beam\cite{Allen&al:PRA:1992}. Also radio beams can be prepared in pure  
OAM states and superpositions thereof such that they contain vortices   
\cite{Thide&al:PRL:2007}.                                               

\begin{figure*}[t]
\centering
\resizebox{.85\textwidth}{!}{%
 \input{fig1.pspdftex}
}
\begin{multicols}{2}
\caption{%
\textbf{Single antenna signal.}\, \textbf{a}, The intensity of
the radio beam in the far zone as obtained in numerical simulations.
\textbf{b}, Experimental results obtained by probing the
radio beam in a plane perpendicular to the beam axis.  The results
exhibit a vortex singularity consistent with the numerical simulations.
}
\label{fig:oneantenna}
\end{multicols}
\end{figure*}
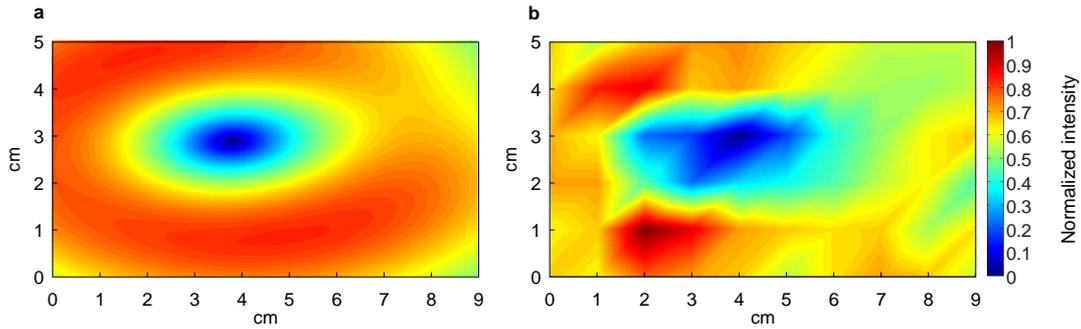

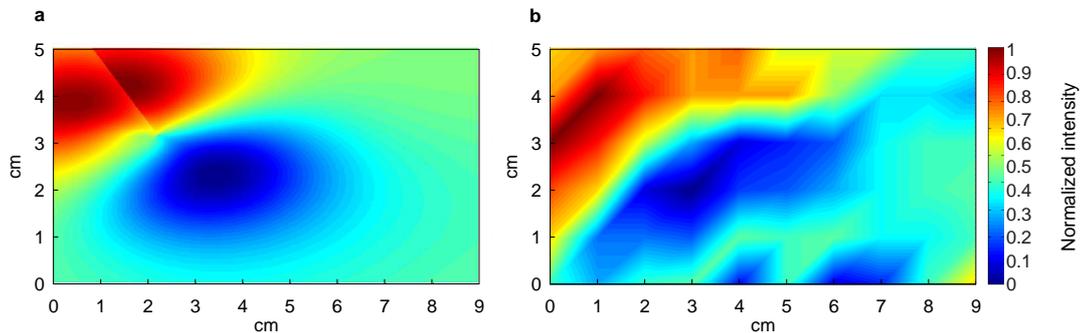
\begin{figure*}[t]
\centering
\resizebox{.85\textwidth}{!}{%
 \input{fig2.pspdftex}
}

\begin{multicols}{2}
\caption{%
\textbf{Sum of two antenna signals.}\,
 These plots show the intensity of the sum of the signal from one       
 antenna at a maximum of the field, and the signal from a    
 second antenna moved to different positions in a plane perpendicular   
 to the beam axis. \textbf{a}, The numerical simulations.
 \textbf{b}, The experimental results. The similarity of the   
 patterns between simulations and experiments shows that the radio      
 beam generated has topological signatures that are consistent with     
 vorticity and OAM.                                                     
}
\label{fig:twoantennas}
\end{multicols}
\end{figure*}

\begin{figure*}[!ht]
\centering
\resizebox{.85\textwidth}{!}{%
 \input{Fig3.pspdftex}
}
\begin{multicols}{2}
\caption{%
 \textbf{Phase map.}\,
 \textbf{a}, Simulated phase map of the radio beam in a far-zone plane 
 perpendicular to the beam axis. The model predicts a $2\pi$ phase step 
 at $-45^\circ$ (the 7 o'clock direction). \textbf{b}, Interpolated phase map,
 based on measurements of the interference of two antenna signals as     
 described above. The low lateral sampling resolution notwithstanding,  
 the map clearly indicates the predicted $2\pi$ phase step across the 7  
 o'clock line.                                                          
}
\label{fig:phase_2D}
\end{multicols}
\end{figure*}
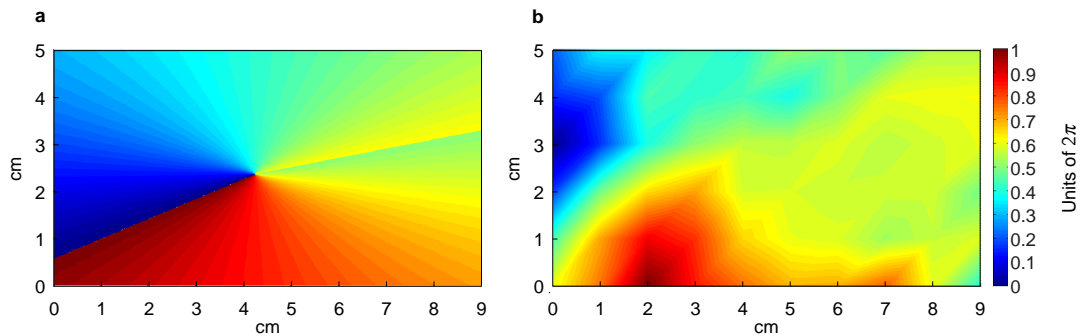

Here we report the experimental confirmation of the generation          
and detection of a radio beam endowed with OAM and harbouring an        
electromagnetic vortex. The results were obtained in controlled         
laboratory experiments by employing the technique of letting a          
radio beam illuminate an approximately spiral-formed          
reflector, realised in $8$ discontinuous steps and sampling the        
reflected radio beam in the far zone by using two identical dipole
antennas.

If $\lambda$ denotes the radio wavelength, and the reflector has        
$N$ discrete jumps and a surface pitch $\eta_\sigma$ (in the            
left-handed sense), the reflected beam will acquire    
the topological charge\cite{Mari&al:OE:2010}                            
\begin{align}
\label{eq:ell}               
 \ell=\frac{2\eta_\sigma}{\lambda}\bigg(\frac{N+1}{N}\bigg)
\end{align}  
For $\eta_\sigma=-6.22$~cm, $\lambda=12.49$~cm ($\nu=2.40$~GHz) and $N=8$ 
as in the experiments reported here, this formula predicts that the     
reflected radio beam had an effective OAM value of $\ell\sim-1.12$.     
According to the formula, a pure $\ell=-1$ OAM eigenstate would have         
been achieved for $\lambda=13.99$~cm ($\nu=2.14$~GHz). However,     
the commercial $2.4$--$2.5$~GHz LAN communications Yagi-Uda antenna     
that was used for generating the primary radio beam did not not work    
satisfactorily at this frequency.                                       

Our findings are summarized in Figures\                                 
\ref{fig:oneantenna}--\ref{fig:phase_2D}. Fig.~\ref{fig:oneantenna}a    
shows the prediction of the beam intensity distribution obtained in a   
numerical simulation, based on equation~(\ref{eq:ell}) for $2.40$~GHz.  
Fig.~\ref{fig:oneantenna}b shows the outcome of the experiment. As      
is clearly seen, the existence of an annularly shaped intensity         
distribution around a central vortex eye, consistent with the existence 
of a vortex in the radio beam, was experimentally confirmed. The field   
intensity is weaker in the 2 o'clock direction, both in the simulation  
and the experiments. This is an effect of the non-integer value of      
$\ell$ and of the corresponding imperfect phase matching over the full  
circle around the vortex eye. Figs.~\ref{fig:twoantennas}a--b      
show, respectively, the numerical and experimental results obtained     
when two receiving antennas were used, one held fixed at a maximum      
field point, and the other sampling different positions in a plane      
transverse to the radio beam axis. In this case the constructive        
and destructive interference of the two signals clearly result in a     
division of the map, along the diagonal, into two regions. Finally,       
we present in Fig.~\ref{fig:phase_2D} an estimate of the comparison     
between the simulated and experimental phase map, interpolated from     
the one- and two-antenna measurements. Despite the unavoidable          
undersampling in the measurements and an ensuing limited accuracy, the  
phase behaviour, particularly the $2\pi$ phase step in the 7 o'clock    
direction, clearly indicates that the beam carries OAM. We notice that  
the symmetry axis of the vortex and the phase singularity are both      
aligned with the mask gap direction of $-45^\circ$ relative to the      
electric field polarization direction, as expected. The conclusion is   
that the experiment produced a radio-frequency vortex, in agreement     
with theory and numerical simulations.                                  

The experimental verification of vorticity and OAM in radio means       
that a new frequency range has become available for fundamental         
as well as applied OAM-based experiments in disciplines ranging         
from relativistic astrophysics\cite{Tamburini&al:NP:2011}               
and nanotechnology and biology\cite{Grier:N:2003}, to                   
wireless communication with high spectral efficiency both               
classically\cite{Gibson&al:OE:2004,Gibson&al:SPIE:2004,
Celechovsky&Bouchal:NJP:2007,Jiang&al:Inproceedings:2009} and           
quantum mechanically\cite{Pors&al:PRL:2008,Barreiro&al:NP:2008}.        
It also opens for the development of new radio                          
and radar probing techniques, including spiral                          
imaging\cite{Thide:PPCF:2007,Molina-Terriza&al:JEOS:2007}. It           
should also be emphasized that certain physical effects and             
observables associated with electromagnetic OAM, for instance           
electromagnetic torque, are stronger for lower frequencies than for     
higher\cite{Carrara:N:1949}.                                            

Perhaps the most striking practical applications of photon vorticity in 
optics and astronomy are super-resolution, where the             
Rayleigh limit can be overcome by at least one order of            
magnitude\cite{Tamburini&al:PRL:2006}, and highly efficient             
coronagraphy, where a contrast of up to $10^{10}$ can be                 
achieved\cite{Swartzlander:OL:2001,Lee&al:PRL:2006,Swartzlander&al:OE:2008,
Anzolin&al:AA:2008,Barbieri&al:EMP:2009,Mawet&al:APJ:2010,
Serabyn&al:N:2010,Mari&al:OE:2010}.
In accordance with the trivial fact that Maxwell's equations are equally 
valid for optical and radio frequencies, our experiments show that the  
new OAM-based techniques developed and employed in optics can indeed    
be adapted to radio. Add to that the fact that electromagnetic orbital  
angular momentum for a large range of frequencies, including radio,     
can be used to reveal ionospheric, magnetospheric,                      
interplanetary, and interstellar plasma                                 
turbulence\cite{Thide:PPCF:2007,Tamburini&al:EPL:2010} as well as       
rotating black holes\cite{Tamburini&al:NP:2011} and one realises        
that we are likely to witness a very rapid development in the space     
sciences.                                                               

\begin{methods}

The experiments were performed in the anechoic antenna chamber at the   
Ångström Laboratory of the Uppsala University, Uppsala, Sweden. The     
chamber is electromagnetically as well as acoustically/vibrationally    
shielded from the outside world. The experiments were performed at      
$2.4$ gigahertz ($12.49$ cm wavelength) with a resonant 7-element       
Yagi-Uda antenna fed from a signal generator with a continuous          
$0.01$ watt signal. The radio beam was reflected off a discrete         
eight-step approximation of a spiral reflector designed for a total     
$2\pi$ phase shift, mimicking an ideal, smooth spiral reflector. The    
reflected, twisted radio beam was probed with two resonant electric     
dipole antennas, on axis in the far zone, $7$ metres ($\sim55\lambda$)     
away from the non-focusing reflector. One of the receiving antennas was 
held at a fixed position whereas the other was moved around to sample   
the signal at different fixed grid points in the one and the same       
plane perpendicular to the radio beam axis. By making a simultaneous    
measurement of the two antenna signals, the absolute amplitude and      
the relative phase of the electric field of the antenna beam could be   
estimated.                                                              

\end{methods}

\small

\begin{acknowledgements}
  B.\,T. acknowledges the financial support from the Swedish Research   
  Council (VR) and the hospitality of the Nordic Institute for          
  Theoretical Physics (NORDITA), Stockholm. F.\,T. and E.\,M.           
  gratefully acknowledge the support from the Vortici e Frequenze       
  group, Orseolo Restauri, CARIPARO Foundation within the 2006 and      
  2008 Program of Excellence, and the kind hospitality of Uppsala       
  University, The Ångström Laboratory antenna chamber was funded by a   
  grant from the Knut and Alice Wallenberg Foundation. We thank Anders  
  Rydberg and Jonas Bothén for technical assistance.                    
\end{acknowledgements}

\end{multicols}

\end{document}

%% file: fig1.pspdftex
\begin{picture}(0,0)%
\includegraphics{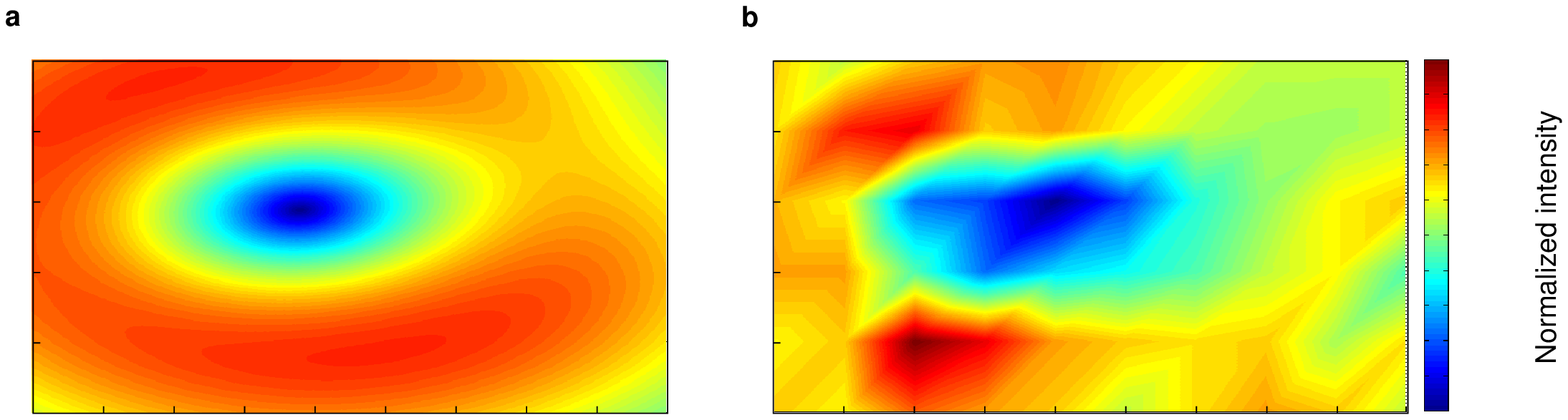}%
\end{picture}%
\setlength{\unitlength}{4144sp}%
\begingroup\makeatletter\ifx\SetFigFont\undefined%
\gdef\SetFigFont#1#2#3#4#5{%
  \reset@font\fontsize{#1}{#2pt}%
  \fontfamily{#3}\fontseries{#4}\fontshape{#5}%
  \selectfont}%
\fi\endgroup%
\begin{picture}(10469,3102)(-225,-2326)
\put(5093,-1925){\makebox(0,0)[rb]{\smash{{\SetFigFont{12}{14.4}{\sfdefault}{\mddefault}{\updefault}{\color[rgb]{0,0,0}0}%
}}}}
\put(5093,-1475){\makebox(0,0)[rb]{\smash{{\SetFigFont{12}{14.4}{\sfdefault}{\mddefault}{\updefault}{\color[rgb]{0,0,0}1}%
}}}}
\put(5093,-1025){\makebox(0,0)[rb]{\smash{{\SetFigFont{12}{14.4}{\sfdefault}{\mddefault}{\updefault}{\color[rgb]{0,0,0}2}%
}}}}
\put(5093,-575){\makebox(0,0)[rb]{\smash{{\SetFigFont{12}{14.4}{\sfdefault}{\mddefault}{\updefault}{\color[rgb]{0,0,0}3}%
}}}}
\put(5093,-125){\makebox(0,0)[rb]{\smash{{\SetFigFont{12}{14.4}{\sfdefault}{\mddefault}{\updefault}{\color[rgb]{0,0,0}4}%
}}}}
\put(5093,325){\makebox(0,0)[rb]{\smash{{\SetFigFont{12}{14.4}{\sfdefault}{\mddefault}{\updefault}{\color[rgb]{0,0,0}5}%
}}}}
\put(5176,-2131){\makebox(0,0)[b]{\smash{{\SetFigFont{12}{14.4}{\sfdefault}{\mddefault}{\updefault}{\color[rgb]{0,0,0}0}%
}}}}
\put(5626,-2131){\makebox(0,0)[b]{\smash{{\SetFigFont{12}{14.4}{\sfdefault}{\mddefault}{\updefault}{\color[rgb]{0,0,0}1}%
}}}}
\put(6076,-2131){\makebox(0,0)[b]{\smash{{\SetFigFont{12}{14.4}{\sfdefault}{\mddefault}{\updefault}{\color[rgb]{0,0,0}2}%
}}}}
\put(6526,-2131){\makebox(0,0)[b]{\smash{{\SetFigFont{12}{14.4}{\sfdefault}{\mddefault}{\updefault}{\color[rgb]{0,0,0}3}%
}}}}
\put(6976,-2131){\makebox(0,0)[b]{\smash{{\SetFigFont{12}{14.4}{\sfdefault}{\mddefault}{\updefault}{\color[rgb]{0,0,0}4}%
}}}}
\put(7426,-2131){\makebox(0,0)[b]{\smash{{\SetFigFont{12}{14.4}{\sfdefault}{\mddefault}{\updefault}{\color[rgb]{0,0,0}5}%
}}}}
\put(7876,-2131){\makebox(0,0)[b]{\smash{{\SetFigFont{12}{14.4}{\sfdefault}{\mddefault}{\updefault}{\color[rgb]{0,0,0}6}%
}}}}
\put(8326,-2131){\makebox(0,0)[b]{\smash{{\SetFigFont{12}{14.4}{\sfdefault}{\mddefault}{\updefault}{\color[rgb]{0,0,0}7}%
}}}}
\put(8776,-2131){\makebox(0,0)[b]{\smash{{\SetFigFont{12}{14.4}{\sfdefault}{\mddefault}{\updefault}{\color[rgb]{0,0,0}8}%
}}}}
\put(9226,-2131){\makebox(0,0)[b]{\smash{{\SetFigFont{12}{14.4}{\sfdefault}{\mddefault}{\updefault}{\color[rgb]{0,0,0}9}%
}}}}
\put(7201,-2311){\makebox(0,0)[b]{\smash{{\SetFigFont{12}{14.4}{\sfdefault}{\mddefault}{\updefault}{\color[rgb]{0,0,0}cm}%
}}}}
\put(4861,-736){\rotatebox{90.0}{\makebox(0,0)[b]{\smash{{\SetFigFont{12}{14.4}{\sfdefault}{\mddefault}{\updefault}{\color[rgb]{0,0,0}cm}%
}}}}}
\put(368,-1925){\makebox(0,0)[rb]{\smash{{\SetFigFont{12}{14.4}{\sfdefault}{\mddefault}{\updefault}{\color[rgb]{0,0,0}0}%
}}}}
\put(368,-1475){\makebox(0,0)[rb]{\smash{{\SetFigFont{12}{14.4}{\sfdefault}{\mddefault}{\updefault}{\color[rgb]{0,0,0}1}%
}}}}
\put(368,-1025){\makebox(0,0)[rb]{\smash{{\SetFigFont{12}{14.4}{\sfdefault}{\mddefault}{\updefault}{\color[rgb]{0,0,0}2}%
}}}}
\put(368,-575){\makebox(0,0)[rb]{\smash{{\SetFigFont{12}{14.4}{\sfdefault}{\mddefault}{\updefault}{\color[rgb]{0,0,0}3}%
}}}}
\put(368,-125){\makebox(0,0)[rb]{\smash{{\SetFigFont{12}{14.4}{\sfdefault}{\mddefault}{\updefault}{\color[rgb]{0,0,0}4}%
}}}}
\put(368,325){\makebox(0,0)[rb]{\smash{{\SetFigFont{12}{14.4}{\sfdefault}{\mddefault}{\updefault}{\color[rgb]{0,0,0}5}%
}}}}
\put(451,-2131){\makebox(0,0)[b]{\smash{{\SetFigFont{12}{14.4}{\sfdefault}{\mddefault}{\updefault}{\color[rgb]{0,0,0}0}%
}}}}
\put(901,-2131){\makebox(0,0)[b]{\smash{{\SetFigFont{12}{14.4}{\sfdefault}{\mddefault}{\updefault}{\color[rgb]{0,0,0}1}%
}}}}
\put(1351,-2131){\makebox(0,0)[b]{\smash{{\SetFigFont{12}{14.4}{\sfdefault}{\mddefault}{\updefault}{\color[rgb]{0,0,0}2}%
}}}}
\put(1801,-2131){\makebox(0,0)[b]{\smash{{\SetFigFont{12}{14.4}{\sfdefault}{\mddefault}{\updefault}{\color[rgb]{0,0,0}3}%
}}}}
\put(2251,-2131){\makebox(0,0)[b]{\smash{{\SetFigFont{12}{14.4}{\sfdefault}{\mddefault}{\updefault}{\color[rgb]{0,0,0}4}%
}}}}
\put(2701,-2131){\makebox(0,0)[b]{\smash{{\SetFigFont{12}{14.4}{\sfdefault}{\mddefault}{\updefault}{\color[rgb]{0,0,0}5}%
}}}}
\put(3151,-2131){\makebox(0,0)[b]{\smash{{\SetFigFont{12}{14.4}{\sfdefault}{\mddefault}{\updefault}{\color[rgb]{0,0,0}6}%
}}}}
\put(3601,-2131){\makebox(0,0)[b]{\smash{{\SetFigFont{12}{14.4}{\sfdefault}{\mddefault}{\updefault}{\color[rgb]{0,0,0}7}%
}}}}
\put(4051,-2131){\makebox(0,0)[b]{\smash{{\SetFigFont{12}{14.4}{\sfdefault}{\mddefault}{\updefault}{\color[rgb]{0,0,0}8}%
}}}}
\put(4501,-2131){\makebox(0,0)[b]{\smash{{\SetFigFont{12}{14.4}{\sfdefault}{\mddefault}{\updefault}{\color[rgb]{0,0,0}9}%
}}}}
\put(2476,-2311){\makebox(0,0)[b]{\smash{{\SetFigFont{12}{14.4}{\sfdefault}{\mddefault}{\updefault}{\color[rgb]{0,0,0}cm}%
}}}}
\put(136,-736){\rotatebox{90.0}{\makebox(0,0)[b]{\smash{{\SetFigFont{12}{14.4}{\sfdefault}{\mddefault}{\updefault}{\color[rgb]{0,0,0}cm}%
}}}}}
\put(9513,-1924){\makebox(0,0)[lb]{\smash{{\SetFigFont{12}{14.4}{\sfdefault}{\mddefault}{\updefault}0}}}}
\put(9513,-1699){\makebox(0,0)[lb]{\smash{{\SetFigFont{12}{14.4}{\sfdefault}{\mddefault}{\updefault}0.1}}}}
\put(9513,-1474){\makebox(0,0)[lb]{\smash{{\SetFigFont{12}{14.4}{\sfdefault}{\mddefault}{\updefault}0.2}}}}
\put(9513,-1249){\makebox(0,0)[lb]{\smash{{\SetFigFont{12}{14.4}{\sfdefault}{\mddefault}{\updefault}0.3}}}}
\put(9513,-1024){\makebox(0,0)[lb]{\smash{{\SetFigFont{12}{14.4}{\sfdefault}{\mddefault}{\updefault}0.4}}}}
\put(9513,-799){\makebox(0,0)[lb]{\smash{{\SetFigFont{12}{14.4}{\sfdefault}{\mddefault}{\updefault}0.5}}}}
\put(9513,-574){\makebox(0,0)[lb]{\smash{{\SetFigFont{12}{14.4}{\sfdefault}{\mddefault}{\updefault}0.6}}}}
\put(9513,-349){\makebox(0,0)[lb]{\smash{{\SetFigFont{12}{14.4}{\sfdefault}{\mddefault}{\updefault}0.7}}}}
\put(9513,-124){\makebox(0,0)[lb]{\smash{{\SetFigFont{12}{14.4}{\sfdefault}{\mddefault}{\updefault}0.8}}}}
\put(9513,101){\makebox(0,0)[lb]{\smash{{\SetFigFont{12}{14.4}{\sfdefault}{\mddefault}{\updefault}0.9}}}}
\put(9513,326){\makebox(0,0)[lb]{\smash{{\SetFigFont{12}{14.4}{\sfdefault}{\mddefault}{\updefault}1}}}}
\end{picture}%

%% file: fig2.pspdftex
\begin{picture}(0,0)%
\includegraphics{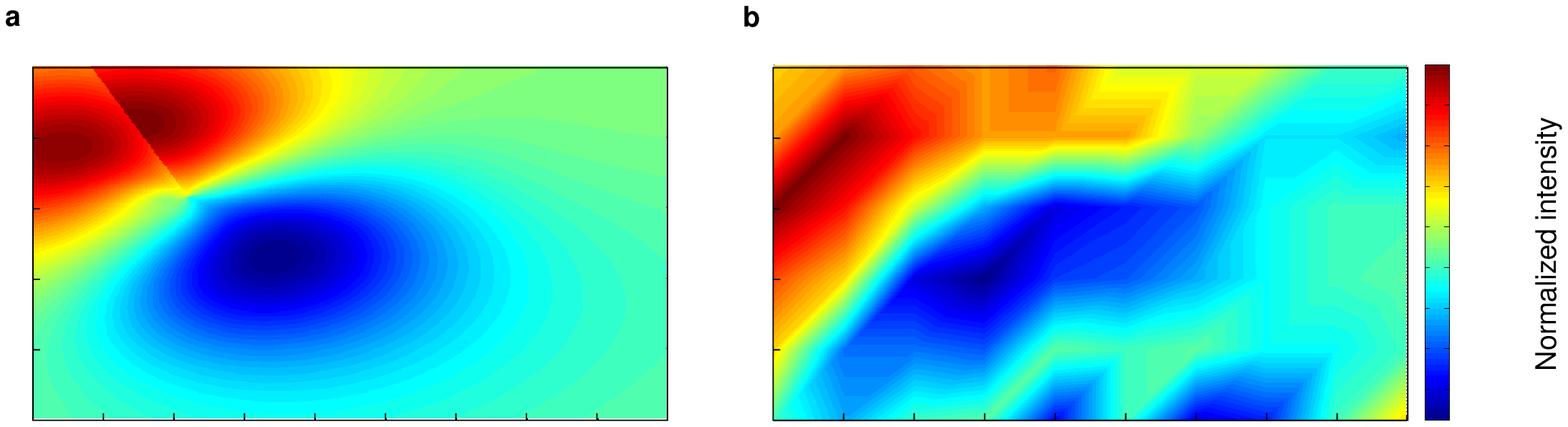}%
\end{picture}%
\setlength{\unitlength}{4144sp}%
\begingroup\makeatletter\ifx\SetFigFont\undefined%
\gdef\SetFigFont#1#2#3#4#5{%
  \reset@font\fontsize{#1}{#2pt}%
  \fontfamily{#3}\fontseries{#4}\fontshape{#5}%
  \selectfont}%
\fi\endgroup%
\begin{picture}(10477,3143)(-233,-2326)
\put(9513,-1924){\makebox(0,0)[lb]{\smash{{\SetFigFont{12}{14.4}{\sfdefault}{\mddefault}{\updefault}0}}}}
\put(9513,-1699){\makebox(0,0)[lb]{\smash{{\SetFigFont{12}{14.4}{\sfdefault}{\mddefault}{\updefault}0.1}}}}
\put(9513,-1474){\makebox(0,0)[lb]{\smash{{\SetFigFont{12}{14.4}{\sfdefault}{\mddefault}{\updefault}0.2}}}}
\put(9513,-1249){\makebox(0,0)[lb]{\smash{{\SetFigFont{12}{14.4}{\sfdefault}{\mddefault}{\updefault}0.3}}}}
\put(9513,-1024){\makebox(0,0)[lb]{\smash{{\SetFigFont{12}{14.4}{\sfdefault}{\mddefault}{\updefault}0.4}}}}
\put(9513,-799){\makebox(0,0)[lb]{\smash{{\SetFigFont{12}{14.4}{\sfdefault}{\mddefault}{\updefault}0.5}}}}
\put(9513,-574){\makebox(0,0)[lb]{\smash{{\SetFigFont{12}{14.4}{\sfdefault}{\mddefault}{\updefault}0.6}}}}
\put(9513,-349){\makebox(0,0)[lb]{\smash{{\SetFigFont{12}{14.4}{\sfdefault}{\mddefault}{\updefault}0.7}}}}
\put(9513,-124){\makebox(0,0)[lb]{\smash{{\SetFigFont{12}{14.4}{\sfdefault}{\mddefault}{\updefault}0.8}}}}
\put(9513,101){\makebox(0,0)[lb]{\smash{{\SetFigFont{12}{14.4}{\sfdefault}{\mddefault}{\updefault}0.9}}}}
\put(9513,326){\makebox(0,0)[lb]{\smash{{\SetFigFont{12}{14.4}{\sfdefault}{\mddefault}{\updefault}1}}}}
\put(368,-1925){\makebox(0,0)[rb]{\smash{{\SetFigFont{12}{14.4}{\sfdefault}{\mddefault}{\updefault}{\color[rgb]{0,0,0}0}%
}}}}
\put(368,-1475){\makebox(0,0)[rb]{\smash{{\SetFigFont{12}{14.4}{\sfdefault}{\mddefault}{\updefault}{\color[rgb]{0,0,0}1}%
}}}}
\put(368,-1025){\makebox(0,0)[rb]{\smash{{\SetFigFont{12}{14.4}{\sfdefault}{\mddefault}{\updefault}{\color[rgb]{0,0,0}2}%
}}}}
\put(368,-575){\makebox(0,0)[rb]{\smash{{\SetFigFont{12}{14.4}{\sfdefault}{\mddefault}{\updefault}{\color[rgb]{0,0,0}3}%
}}}}
\put(368,-125){\makebox(0,0)[rb]{\smash{{\SetFigFont{12}{14.4}{\sfdefault}{\mddefault}{\updefault}{\color[rgb]{0,0,0}4}%
}}}}
\put(368,325){\makebox(0,0)[rb]{\smash{{\SetFigFont{12}{14.4}{\sfdefault}{\mddefault}{\updefault}{\color[rgb]{0,0,0}5}%
}}}}
\put(451,-2131){\makebox(0,0)[b]{\smash{{\SetFigFont{12}{14.4}{\sfdefault}{\mddefault}{\updefault}{\color[rgb]{0,0,0}0}%
}}}}
\put(901,-2131){\makebox(0,0)[b]{\smash{{\SetFigFont{12}{14.4}{\sfdefault}{\mddefault}{\updefault}{\color[rgb]{0,0,0}1}%
}}}}
\put(1351,-2131){\makebox(0,0)[b]{\smash{{\SetFigFont{12}{14.4}{\sfdefault}{\mddefault}{\updefault}{\color[rgb]{0,0,0}2}%
}}}}
\put(1801,-2131){\makebox(0,0)[b]{\smash{{\SetFigFont{12}{14.4}{\sfdefault}{\mddefault}{\updefault}{\color[rgb]{0,0,0}3}%
}}}}
\put(2251,-2131){\makebox(0,0)[b]{\smash{{\SetFigFont{12}{14.4}{\sfdefault}{\mddefault}{\updefault}{\color[rgb]{0,0,0}4}%
}}}}
\put(2701,-2131){\makebox(0,0)[b]{\smash{{\SetFigFont{12}{14.4}{\sfdefault}{\mddefault}{\updefault}{\color[rgb]{0,0,0}5}%
}}}}
\put(3151,-2131){\makebox(0,0)[b]{\smash{{\SetFigFont{12}{14.4}{\sfdefault}{\mddefault}{\updefault}{\color[rgb]{0,0,0}6}%
}}}}
\put(3601,-2131){\makebox(0,0)[b]{\smash{{\SetFigFont{12}{14.4}{\sfdefault}{\mddefault}{\updefault}{\color[rgb]{0,0,0}7}%
}}}}
\put(4051,-2131){\makebox(0,0)[b]{\smash{{\SetFigFont{12}{14.4}{\sfdefault}{\mddefault}{\updefault}{\color[rgb]{0,0,0}8}%
}}}}
\put(4501,-2131){\makebox(0,0)[b]{\smash{{\SetFigFont{12}{14.4}{\sfdefault}{\mddefault}{\updefault}{\color[rgb]{0,0,0}9}%
}}}}
\put(2476,-2311){\makebox(0,0)[b]{\smash{{\SetFigFont{12}{14.4}{\sfdefault}{\mddefault}{\updefault}{\color[rgb]{0,0,0}cm}%
}}}}
\put(136,-736){\rotatebox{90.0}{\makebox(0,0)[b]{\smash{{\SetFigFont{12}{14.4}{\sfdefault}{\mddefault}{\updefault}{\color[rgb]{0,0,0}cm}%
}}}}}
\put(5093,-1925){\makebox(0,0)[rb]{\smash{{\SetFigFont{12}{14.4}{\sfdefault}{\mddefault}{\updefault}{\color[rgb]{0,0,0}0}%
}}}}
\put(5093,-1475){\makebox(0,0)[rb]{\smash{{\SetFigFont{12}{14.4}{\sfdefault}{\mddefault}{\updefault}{\color[rgb]{0,0,0}1}%
}}}}
\put(5093,-1025){\makebox(0,0)[rb]{\smash{{\SetFigFont{12}{14.4}{\sfdefault}{\mddefault}{\updefault}{\color[rgb]{0,0,0}2}%
}}}}
\put(5093,-575){\makebox(0,0)[rb]{\smash{{\SetFigFont{12}{14.4}{\sfdefault}{\mddefault}{\updefault}{\color[rgb]{0,0,0}3}%
}}}}
\put(5093,-125){\makebox(0,0)[rb]{\smash{{\SetFigFont{12}{14.4}{\sfdefault}{\mddefault}{\updefault}{\color[rgb]{0,0,0}4}%
}}}}
\put(5093,325){\makebox(0,0)[rb]{\smash{{\SetFigFont{12}{14.4}{\sfdefault}{\mddefault}{\updefault}{\color[rgb]{0,0,0}5}%
}}}}
\put(5176,-2131){\makebox(0,0)[b]{\smash{{\SetFigFont{12}{14.4}{\sfdefault}{\mddefault}{\updefault}{\color[rgb]{0,0,0}0}%
}}}}
\put(5626,-2131){\makebox(0,0)[b]{\smash{{\SetFigFont{12}{14.4}{\sfdefault}{\mddefault}{\updefault}{\color[rgb]{0,0,0}1}%
}}}}
\put(6076,-2131){\makebox(0,0)[b]{\smash{{\SetFigFont{12}{14.4}{\sfdefault}{\mddefault}{\updefault}{\color[rgb]{0,0,0}2}%
}}}}
\put(6526,-2131){\makebox(0,0)[b]{\smash{{\SetFigFont{12}{14.4}{\sfdefault}{\mddefault}{\updefault}{\color[rgb]{0,0,0}3}%
}}}}
\put(6976,-2131){\makebox(0,0)[b]{\smash{{\SetFigFont{12}{14.4}{\sfdefault}{\mddefault}{\updefault}{\color[rgb]{0,0,0}4}%
}}}}
\put(7426,-2131){\makebox(0,0)[b]{\smash{{\SetFigFont{12}{14.4}{\sfdefault}{\mddefault}{\updefault}{\color[rgb]{0,0,0}5}%
}}}}
\put(7876,-2131){\makebox(0,0)[b]{\smash{{\SetFigFont{12}{14.4}{\sfdefault}{\mddefault}{\updefault}{\color[rgb]{0,0,0}6}%
}}}}
\put(8326,-2131){\makebox(0,0)[b]{\smash{{\SetFigFont{12}{14.4}{\sfdefault}{\mddefault}{\updefault}{\color[rgb]{0,0,0}7}%
}}}}
\put(8776,-2131){\makebox(0,0)[b]{\smash{{\SetFigFont{12}{14.4}{\sfdefault}{\mddefault}{\updefault}{\color[rgb]{0,0,0}8}%
}}}}
\put(9226,-2131){\makebox(0,0)[b]{\smash{{\SetFigFont{12}{14.4}{\sfdefault}{\mddefault}{\updefault}{\color[rgb]{0,0,0}9}%
}}}}
\put(7201,-2311){\makebox(0,0)[b]{\smash{{\SetFigFont{12}{14.4}{\sfdefault}{\mddefault}{\updefault}{\color[rgb]{0,0,0}cm}%
}}}}
\put(4861,-736){\rotatebox{90.0}{\makebox(0,0)[b]{\smash{{\SetFigFont{12}{14.4}{\sfdefault}{\mddefault}{\updefault}{\color[rgb]{0,0,0}cm}%
}}}}}
\end{picture}%

%% file: Fig3.pspdftex
\begin{picture}(0,0)%
\includegraphics{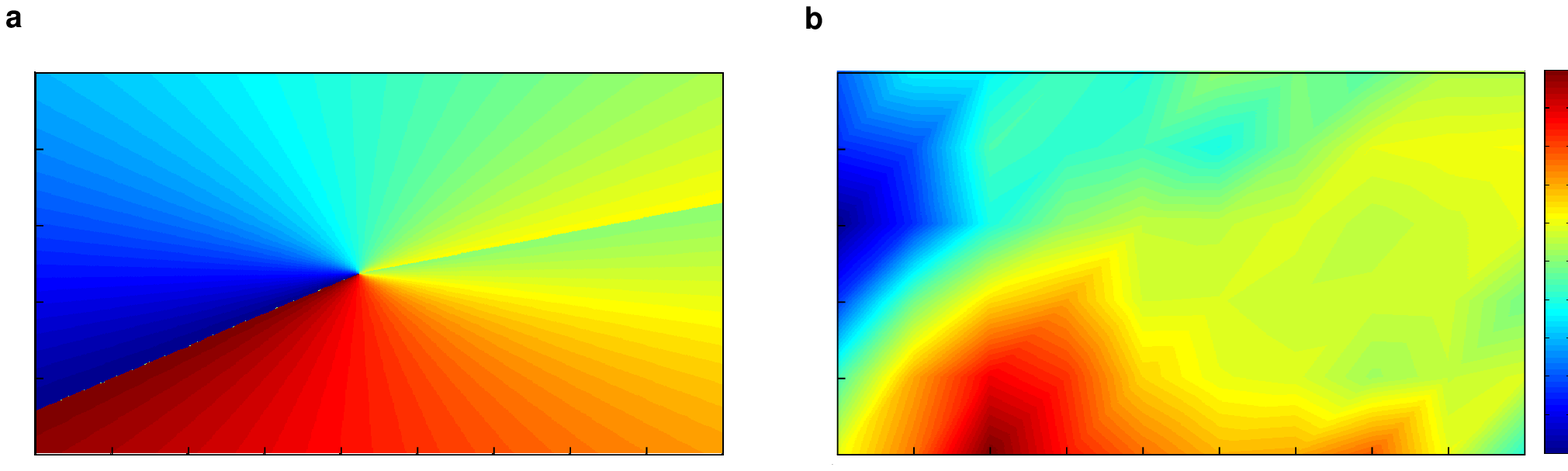}%
\end{picture}%
\setlength{\unitlength}{4144sp}%
\begingroup\makeatletter\ifx\SetFigFont\undefined%
\gdef\SetFigFont#1#2#3#4#5{%
  \reset@font\fontsize{#1}{#2pt}%
  \fontfamily{#3}\fontseries{#4}\fontshape{#5}%
  \selectfont}%
\fi\endgroup%
\begin{picture}(10435,3143)(-236,-2326)
\put(368,-1925){\makebox(0,0)[rb]{\smash{{\SetFigFont{12}{14.4}{\sfdefault}{\mddefault}{\updefault}{\color[rgb]{0,0,0}0}%
}}}}
\put(368,-1475){\makebox(0,0)[rb]{\smash{{\SetFigFont{12}{14.4}{\sfdefault}{\mddefault}{\updefault}{\color[rgb]{0,0,0}1}%
}}}}
\put(368,-1025){\makebox(0,0)[rb]{\smash{{\SetFigFont{12}{14.4}{\sfdefault}{\mddefault}{\updefault}{\color[rgb]{0,0,0}2}%
}}}}
\put(368,-575){\makebox(0,0)[rb]{\smash{{\SetFigFont{12}{14.4}{\sfdefault}{\mddefault}{\updefault}{\color[rgb]{0,0,0}3}%
}}}}
\put(368,-125){\makebox(0,0)[rb]{\smash{{\SetFigFont{12}{14.4}{\sfdefault}{\mddefault}{\updefault}{\color[rgb]{0,0,0}4}%
}}}}
\put(368,325){\makebox(0,0)[rb]{\smash{{\SetFigFont{12}{14.4}{\sfdefault}{\mddefault}{\updefault}{\color[rgb]{0,0,0}5}%
}}}}
\put(451,-2131){\makebox(0,0)[b]{\smash{{\SetFigFont{12}{14.4}{\sfdefault}{\mddefault}{\updefault}{\color[rgb]{0,0,0}0}%
}}}}
\put(901,-2131){\makebox(0,0)[b]{\smash{{\SetFigFont{12}{14.4}{\sfdefault}{\mddefault}{\updefault}{\color[rgb]{0,0,0}1}%
}}}}
\put(1351,-2131){\makebox(0,0)[b]{\smash{{\SetFigFont{12}{14.4}{\sfdefault}{\mddefault}{\updefault}{\color[rgb]{0,0,0}2}%
}}}}
\put(1801,-2131){\makebox(0,0)[b]{\smash{{\SetFigFont{12}{14.4}{\sfdefault}{\mddefault}{\updefault}{\color[rgb]{0,0,0}3}%
}}}}
\put(2251,-2131){\makebox(0,0)[b]{\smash{{\SetFigFont{12}{14.4}{\sfdefault}{\mddefault}{\updefault}{\color[rgb]{0,0,0}4}%
}}}}
\put(2701,-2131){\makebox(0,0)[b]{\smash{{\SetFigFont{12}{14.4}{\sfdefault}{\mddefault}{\updefault}{\color[rgb]{0,0,0}5}%
}}}}
\put(3151,-2131){\makebox(0,0)[b]{\smash{{\SetFigFont{12}{14.4}{\sfdefault}{\mddefault}{\updefault}{\color[rgb]{0,0,0}6}%
}}}}
\put(3601,-2131){\makebox(0,0)[b]{\smash{{\SetFigFont{12}{14.4}{\sfdefault}{\mddefault}{\updefault}{\color[rgb]{0,0,0}7}%
}}}}
\put(4051,-2131){\makebox(0,0)[b]{\smash{{\SetFigFont{12}{14.4}{\sfdefault}{\mddefault}{\updefault}{\color[rgb]{0,0,0}8}%
}}}}
\put(4501,-2131){\makebox(0,0)[b]{\smash{{\SetFigFont{12}{14.4}{\sfdefault}{\mddefault}{\updefault}{\color[rgb]{0,0,0}9}%
}}}}
\put(2476,-2311){\makebox(0,0)[b]{\smash{{\SetFigFont{12}{14.4}{\sfdefault}{\mddefault}{\updefault}{\color[rgb]{0,0,0}cm}%
}}}}
\put(136,-736){\rotatebox{90.0}{\makebox(0,0)[b]{\smash{{\SetFigFont{12}{14.4}{\sfdefault}{\mddefault}{\updefault}{\color[rgb]{0,0,0}cm}%
}}}}}
\put(5093,-1925){\makebox(0,0)[rb]{\smash{{\SetFigFont{12}{14.4}{\sfdefault}{\mddefault}{\updefault}{\color[rgb]{0,0,0}0}%
}}}}
\put(5093,-1475){\makebox(0,0)[rb]{\smash{{\SetFigFont{12}{14.4}{\sfdefault}{\mddefault}{\updefault}{\color[rgb]{0,0,0}1}%
}}}}
\put(5093,-1025){\makebox(0,0)[rb]{\smash{{\SetFigFont{12}{14.4}{\sfdefault}{\mddefault}{\updefault}{\color[rgb]{0,0,0}2}%
}}}}
\put(5093,-575){\makebox(0,0)[rb]{\smash{{\SetFigFont{12}{14.4}{\sfdefault}{\mddefault}{\updefault}{\color[rgb]{0,0,0}3}%
}}}}
\put(5093,-125){\makebox(0,0)[rb]{\smash{{\SetFigFont{12}{14.4}{\sfdefault}{\mddefault}{\updefault}{\color[rgb]{0,0,0}4}%
}}}}
\put(5093,325){\makebox(0,0)[rb]{\smash{{\SetFigFont{12}{14.4}{\sfdefault}{\mddefault}{\updefault}{\color[rgb]{0,0,0}5}%
}}}}
\put(5176,-2131){\makebox(0,0)[b]{\smash{{\SetFigFont{12}{14.4}{\sfdefault}{\mddefault}{\updefault}{\color[rgb]{0,0,0}0}%
}}}}
\put(5626,-2131){\makebox(0,0)[b]{\smash{{\SetFigFont{12}{14.4}{\sfdefault}{\mddefault}{\updefault}{\color[rgb]{0,0,0}1}%
}}}}
\put(6076,-2131){\makebox(0,0)[b]{\smash{{\SetFigFont{12}{14.4}{\sfdefault}{\mddefault}{\updefault}{\color[rgb]{0,0,0}2}%
}}}}
\put(6526,-2131){\makebox(0,0)[b]{\smash{{\SetFigFont{12}{14.4}{\sfdefault}{\mddefault}{\updefault}{\color[rgb]{0,0,0}3}%
}}}}
\put(6976,-2131){\makebox(0,0)[b]{\smash{{\SetFigFont{12}{14.4}{\sfdefault}{\mddefault}{\updefault}{\color[rgb]{0,0,0}4}%
}}}}
\put(7426,-2131){\makebox(0,0)[b]{\smash{{\SetFigFont{12}{14.4}{\sfdefault}{\mddefault}{\updefault}{\color[rgb]{0,0,0}5}%
}}}}
\put(7876,-2131){\makebox(0,0)[b]{\smash{{\SetFigFont{12}{14.4}{\sfdefault}{\mddefault}{\updefault}{\color[rgb]{0,0,0}6}%
}}}}
\put(8326,-2131){\makebox(0,0)[b]{\smash{{\SetFigFont{12}{14.4}{\sfdefault}{\mddefault}{\updefault}{\color[rgb]{0,0,0}7}%
}}}}
\put(8776,-2131){\makebox(0,0)[b]{\smash{{\SetFigFont{12}{14.4}{\sfdefault}{\mddefault}{\updefault}{\color[rgb]{0,0,0}8}%
}}}}
\put(9226,-2131){\makebox(0,0)[b]{\smash{{\SetFigFont{12}{14.4}{\sfdefault}{\mddefault}{\updefault}{\color[rgb]{0,0,0}9}%
}}}}
\put(7201,-2311){\makebox(0,0)[b]{\smash{{\SetFigFont{12}{14.4}{\sfdefault}{\mddefault}{\updefault}{\color[rgb]{0,0,0}cm}%
}}}}
\put(4861,-736){\rotatebox{90.0}{\makebox(0,0)[b]{\smash{{\SetFigFont{12}{14.4}{\sfdefault}{\mddefault}{\updefault}{\color[rgb]{0,0,0}cm}%
}}}}}
\put(9513,-1924){\makebox(0,0)[lb]{\smash{{\SetFigFont{12}{14.4}{\sfdefault}{\mddefault}{\updefault}0}}}}
\put(9513,-1699){\makebox(0,0)[lb]{\smash{{\SetFigFont{12}{14.4}{\sfdefault}{\mddefault}{\updefault}0.1}}}}
\put(9513,-1474){\makebox(0,0)[lb]{\smash{{\SetFigFont{12}{14.4}{\sfdefault}{\mddefault}{\updefault}0.2}}}}
\put(9513,-1249){\makebox(0,0)[lb]{\smash{{\SetFigFont{12}{14.4}{\sfdefault}{\mddefault}{\updefault}0.3}}}}
\put(9513,-1024){\makebox(0,0)[lb]{\smash{{\SetFigFont{12}{14.4}{\sfdefault}{\mddefault}{\updefault}0.4}}}}
\put(9513,-799){\makebox(0,0)[lb]{\smash{{\SetFigFont{12}{14.4}{\sfdefault}{\mddefault}{\updefault}0.5}}}}
\put(9513,-574){\makebox(0,0)[lb]{\smash{{\SetFigFont{12}{14.4}{\sfdefault}{\mddefault}{\updefault}0.6}}}}
\put(9513,-349){\makebox(0,0)[lb]{\smash{{\SetFigFont{12}{14.4}{\sfdefault}{\mddefault}{\updefault}0.7}}}}
\put(9513,-124){\makebox(0,0)[lb]{\smash{{\SetFigFont{12}{14.4}{\sfdefault}{\mddefault}{\updefault}0.8}}}}
\put(9513,101){\makebox(0,0)[lb]{\smash{{\SetFigFont{12}{14.4}{\sfdefault}{\mddefault}{\updefault}0.9}}}}
\put(9513,326){\makebox(0,0)[lb]{\smash{{\SetFigFont{12}{14.4}{\sfdefault}{\mddefault}{\updefault}1}}}}
\put(10126,-736){\rotatebox{90.0}{\makebox(0,0)[b]{\smash{{\SetFigFont{12}{14.4}{\sfdefault}{\mddefault}{\updefault}{\color[rgb]{0,0,0}\sffamily Units of $\mathsf{2\pi}$}%
}}}}}
\end{picture}%